\documentclass[twocolumn,showpacs,preprintnumbers,amsmath,amssymb,prb,aps]{revtex4}

\usepackage{epsfig}
\usepackage{graphicx}
\usepackage{dcolumn}
\usepackage{bm}

\begin{document}

\title{Electronic structure of PrCoO$_3$ and its temperature evolution}

\author{S.K. Pandey}
\author{Swapnil Patil}
\author{V.R.R. Medicherla}
\author{R.S. Singh}
\author{Kalobaran Maiti}
\altaffiliation{Electronic mail: kbmaiti@tifr.res.in}

\affiliation{Department of Condensed Matter Physics and Materials
Science, Tata Institute of Fundamental Research, Homi Bhabha Road,
Colaba, Mumbai - 400 005, INDIA}

\date{\today}

\begin{abstract}

We investigate the detailed electronic structure of PrCoO$_3$ and
its temperature evolution using state-of-the-art photoemission
spectroscopy and {\em ab initio} band structure calculations. We
observe that in addition to the correlation effect, spin-orbit
interaction plays an important role in determining the electronic
properties of this system. Pr 4$f$ states are found to be strongly
hybridized with the O 2$p$ and Co 3$d$ valence electronic states,
and thus influences the electronic properties significantly. The
calculated results corresponding to the intermediate spin state of
Co provide a good description of the experimental spectra at 300 K.
The decrease in temperature from 300 K leads to a gradual
enhancement of the low spin state contributions in the electronic
structure. The temperature evolution of the band gap is found to be
consistent with the transport data.

\end{abstract}


\maketitle

\section{Introduction}

Cobaltates with general formula ACoO$_{3}$ (A $\equiv$ rare-earth
ions) forms an interesting class of compounds in the perovskite
family.\cite{imada} The primary concerns in these systems is the
understanding of the temperature induced spin state transition.
Extensive effort has been put forward both theoretically and
experimentally to understand this effect. The ground state of these
compounds are believed to be nonmagnetic (spin $S$ = 0) insulator
having Co$^{3+}$ ion in low spin (LS) configuration with fully
filled Co $t_{2g}$ orbitals. Increase in temperature leads to
non-magnetic insulator to paramagnetic metal
transitions.\cite{heikes,bhide,yamaguchi1,yamaguchi2,tsubouchi} It
was suggested that such transitions occurs due to thermally driven
spin state transition of Co$^{3+}$
ions.\cite{goodenough,raccah,korotin} The major controversy in these
studies involves the identification of the spin state in the
intermediate temperatures range. Some
studies\cite{korotin,kobayashi,zobel,radaelli,maris,plakhty,phelan,sudheendra,yan,klie}
suggest that the spin state of Co is an intermediate spin (IS
$\Rightarrow$
$t_{2g\uparrow}^{3}e_{g\uparrow}^{1}t_{2g\downarrow}^{2}$, S = 1)
state, while the others suggest a mixed low spin and high spin (HS
$\Rightarrow$
$t_{2g\uparrow}^{3}e_{g\uparrow}^{2}t_{2g\downarrow}^{1}$, S = 2)
state.\cite{goodenough,raccah,abbate,barman,knizek,haverkort,podlesnyak}
Despite numerous investigations on this issue, the controversy in
the spin state transition is still unresolved.

Most of the studies mentioned above have focussed on the spin state
transitions in LaCoO$_{3}$, where the crystal field splitting is
close to the exchange interaction strength. If we replace La by Pr,
the smaller ionic radius of Pr compared to La provides a chemical
pressure on the system. It is observed that the crystal structure of
PrCoO$_3$ is different from that of LaCoO$_3$. The powder
diffraction work has shown that the crystal structure of LaCoO$_{3}$
and PrCoO$_{3}$ are rhombohedral and orthorhombic,
respectively.\cite{tsubouchi,radaelli} The average Co-O bond length
of PrCoO$_{3}$ is less than that of LaCoO$_{3}$ and hence the
crystal field splitting is expected to be enhanced in PrCoO$_3$. The
Co-O-Co bond angle is smaller in PrCoO$_3$ compared to that in
LaCoO$_3$. In addition, 4$f$ electrons corresponding to Pr atoms and
the hybridization of the 4$f$ states with the valence electronic
states will play an important role in determining the magnetic
properties in this system.

The transport and optical conductivity measurements exhibit a larger
band gap and higher insulator to metal transition temperature in
PrCoO$_3$ compared to that in LaCoO$_3$.\cite{yamaguchi1} It was
suggested that the decrease in valence band width due to the change
in Co-O-Co bond angle leads to such effect. The temperature
dependent infrared spectroscopy data of PrCoO$_{3}$ have shown the
LS to IS state transition around 220 K.\cite{sudheendra} The
susceptibility data of PrCoO$_{3}$ obtained after subtracting the
contributions from Pr$^{3+}$ ions have shown increased population of
the IS state around 200 K.\cite{yan} Thus, PrCoO$_3$ is an ideal
system to throw some light in understanding the interplay between
crystal field splitting and exchange coupling strength in the spin
state transition in cobaltates.

In the present work, we investigate the temperature evolution of
spin state of Co ion in PrCoO$_3$ using state-of-the-art
photoemission spectroscopy and {\em ab initio} band structure
calculations. The experimental valence band spectra exhibit
signature of gradual increase in band gap with the decrease in
temperature. The comparison of the experimental and calculated
results for the valence band indicates that the Co ion presumably
possesses an IS state configuration in the temperature range,
150-300 K. This is also manifested in the shallow core level
spectra. Spin orbit coupling (SOC) plays an important role in
determining the electronic structure in this system.

\section{Experimental and computational details}

PrCoO$_{3}$ was prepared in the polycrystalline form by combustion
method. \cite{pandey2} Nitrates of Pr and Co were taken in
appropriate amount and mixed in double distilled water. In this
mixture, 2 moles of glycine per 1 mole of metal cation was added and
stirred until all compounds dissolved in the water. The resulting
solution was heated slowly at temperature around 200 $^{0}$C till
all the water evaporated. The precursor thus formed catches fire on
its own making the powder of the compound. The hard pallets of this
powder were formed and heated at 1200 $^{0}$C for one day. The
sample was characterized by $x$-ray powder diffraction and
resistivity measurements. The powder diffraction data did not show
any impurity peak; all the peaks were well fitted with orthorhombic
phase (space group Pbnm) using Rietveld refinement technique. The
lattice parameters obtained from the fitting match well with those
reported in the literature. Resistivity data are also consistent
with the available data in the
literature.\cite{yamaguchi1,tsubouchi}

The photoemission spectra were recorded using a spectrometer
equipped with monochromatic sources such as $x$-ray source, Al
$K\alpha$ (1486.6 eV), ultraviolet sources, He {\scriptsize II}
(40.8 eV) and He {\scriptsize I} (21.2 eV). The electron detection
was carried out by a Gammadata Scienta analyzer, SES2002. The energy
resolution for the $x$-ray photoemission (XP) was set to 0.3 eV for
the valence band and 0.6 eV for the core level spectra. The energy
resolution for the He {\scriptsize II} and He {\scriptsize I}
spectra were fixed to 4.2 meV and 1.4 meV, respectively. The base
pressure during the measurements was about 4$\times$10$^{-11}$ Torr.
We have used a He-cryostat, LT-3M from Advanced Research Systems for
the measurements at different temperatures.

The sample was cleaned {\em in situ} by scraping the sample surface
using a diamond file. The cleanliness of the sample was ascertained
by tracking the sharpness of O 1$s$ peak and absence of C 1$s$ peak.
The O 1$s$ spectra recorded at 150 K and 300 K are shown in Fig. 1.
The contribution of impurity peak around 531 eV binding energy is
negligible ($<$~3\%) indicating the good quality of the sample. The
Fermi level was aligned by recording the valence band spectrum of an
Ag foil mounted on the same sample holder.

The GGA (generalized gradient approximation) and GGA+$U$ ($U$ = the
electron correlation strength) calculations were carried out using
LMTART 6.61.\cite{savrasov} For calculating charge density,
full-potential linearized Muffin-Tin orbital method working in plane
wave representation was employed. In the calculation, we have used
the Muffin-Tin radii of 3.43, 2.019, and 1.688 a.u. for Pr, Co and
O, respectively. The charge density and effective potential were
expanded in spherical harmonics up to $l$ = 6 inside the sphere and
in a Fourier series in the interstitial region. The initial basis
set included 6$s$, 5$p$, 5$d$, and 4$f$ valence, and 5$s$ semicore
orbitals of Pr; 4$s$, 4$p$, and 3$d$ valence, and 3$p$ semicore
orbitals of Co, and 2$s$ and 2$p$ valence orbitals of O. The
exchange correlation functional of the density functional theory was
taken after H.S. Vosko {\em et al}.\cite{vosko} and GGA was
implemented using J.P. Perdew {\em et al.}
prescription.\cite{perdew} The effect of spin-orbit coupling (SOC)
was also considered in the calculations.

In the GGA+$U$ calculations the Hubbard $U$ and exchange $J$ are
considered as parameters. We have taken $U$ = 3.5 eV for Co 3$d$ and
Pr 4$f$ electrons and $J$ = 1.0 eV for Co 3d electrons. We have not
considered the value of $J$ for Pr 4$f$ electrons exclusively in the
calculations. It is estimated under GGA formulation. The values of
$U$ and $J$ for 3$d$ electrons are consistent with our previous
studies.\cite{pandey1,pandey3} These calculations were performed by
taking LS, IS and HS configurations, which correspond to
($t_{2g\uparrow}^3 t_{2g\downarrow}^3$), ($t_{2g\uparrow}^3
t_{2g\downarrow}^2 e_{g\uparrow}^1$) and ($t_{2g\uparrow}^3
t_{2g\downarrow}^1 e_{g\uparrow}^2$) electronic configurations,
respectively, as initial input. Self-consistency was achieved by
demanding the convergence of the total energy to be smaller than
10$^{-4}$ Ryd/cell. Final orbital occupancies for Pr 4$f$, Co
$t_{2g}$ and $e_g$ states were obtained from self-consistent GGA+$U$
calculations for different initial state configurations. (8, 8, 6)
divisions of the Brillouin zone along three directions for the
tetrahedron integration were used to calculate the density of states
(DOS).

\section{Results and discussions}

Co $t_{2g}$, Co $e_{g}$ and O 2$p$ partial density of states (PDOS)
obtained from GGA calculations are plotted in Fig. 2(a) and PDOS of
Pr 4$f$ in Fig. 2(b). It is evident from the figures that there are
finite PDOS of Co $e_{g}$, O 2$p$ and Pr 4$f$ at the Fermi level
indicating the metallic ground state which is contrary to the
experimentally observed insulating ground state. The PDOS below the
Fermi level can be divided into three regions: (i) region A up to
-1.6 eV from the Fermi level, (ii) region B from -1.6 eV to -3.4 eV
and (iii) region C below -3.4 eV. Region A has dominating Co 3$d$
character along with small O 2$p$ contributions having similar
energy distribution of the PDOS. This feature is attributed to the
antibonding states. The Co 3$d$ states having $t_{2g}$ symmetry
appear between -1.6 eV to -0.2 eV and the contribution of Co 3$d$
states having $e_{g}$ symmetry appear between -0.5 to 2.5 eV. Region
B is essentially contributed by non-bonding O 2$p$ states. The
bonding states having $t_{2g}$ and $e_g$ symmetry contribute in
region C with dominant contribution from O 2$p$ electronic states.
The contribution of Pr 4$f$ PDOS appears in the narrow region from
-0.1 eV to 0.4 eV indicating the highly localized nature of Pr 4$f$
electrons.

In order to investigate the role of SOC in the electronic structure
in the limit of generalized gradient approximations, we have
calculated the density of states including SOC. The calculated PDOS
of Co $t_{2g}$, Co $e_{g}$ and O 2$p$ states are shown in Fig. 2(c)
and that of Pr 4$f$ states in Fig. 2(d). It is clear from the figure
that inclusion of SOC has negligible influence on the Co 3$d$ and O
2$p$ PDOS. However, SOC drastically modifies the Pr 4$f$ PDOS as
evident from Fig. 2(d). It splits into two well separated regions.
Region below 0.25 eV is identified to be Pr 4$f_{5/2}$ states and
region above 0.25 eV appears due to Pr 4$f_{7/2}$ contributions. The
spin-orbit splitting of Pr 4$f$ states is found to be about 0.4 eV.

It is clear from Fig. 2 that the DOS at the Fermi level,
$\epsilon_F$ is finite in both the cases (with SOC and without SOC);
an indication of a metallic ground state. Thus, the consideration of
electron correlation is necessary to capture the insulating ground
state observed experimentally.

The inclusion of on-site Coulomb correlation on Co 3$d$ electrons
($U_{dd}$ = 3.5 eV) under GGA+$\emph{U}$ formulation modifies the
positions and distributions of Co 3$d$ and O 2$p$ PDOS as evident in
Fig. 3(a). The most significant effect of $U_{dd}$ is observed in
the antibonding Co $t_{2g}$ states (the feature A in the figure),
which shifts to lower energy by about 0.8 eV. The contribution of
the O 2p states enhances significantly in this energy range. As
expected $U_{dd}$ does not affect Pr 4$f$ bands. Interestingly, Co
$e_{g}$ states have finite contribution at the Fermi level, which is
not expected as the inclusion of $U_{dd}$ of 3.5 eV was sufficient
to create a hard gap of about 0.22 eV in LaCoO$_{3}$.\cite{pandey1}
This indicates that Pr 4$f$ states strongly hybridize with the O
2$p$ and Co $e_{g}$ orbitals in the vicinity of the Fermi level.
This is also manifested by the shape of the energy distribution of
the DOS in this energy region. Still, there is finite DOS at
$\epsilon_F$ characterizing the system to be metallic.

In order to capture the correct ground state of PrCoO$_3$, we have
included the on-site Coulomb correlation among Pr 4$f$ electrons
($U_{ff}$). Since, the $f$ electrons are more localized than the $d$
electrons, the electron correlation strength among $f$ electrons is
expected to be larger than that for $d$ electrons. In our
calculations, we have considered $U_{ff}$ to be similar to $U_{dd}$
(= 3.5 eV) so that it corresponds to the lower limit in $U_{ff}$. In
Fig. 3(c), we show the Co $t_{2g}$, Co $e_{g}$ and O 2$p$ PDOS. Pr
4$f$ PDOS obtained are shown in Fig. 3(d). Pr 4$f$ PDOS exhibit a
large splitting leading to the formation of lower and upper Hubbard
bands representing the localized density of states. The energy gap
between these bands is as large as 2.2 eV. Interestingly, such
splitting introduces significant modification in the Co 3$d$ and O
2$p$ DOS. The 4$f$ symmetry adapted Co 3$d$ and O 2$p$ states also
appear in the same energy range as that of the 4$f$ bands and a band
gap of 0.27 eV appears between the $t_{2g}$ and $e_g$ states
characterizing the system to be an insulator.

In addition, the inclusion of $U_{ff}$ introduces a large shift of
the features A, B and C towards the Fermi level along with a
significant change in the energy separation among themselves. While
the feature A is predominantly contributed by Co 3$d$ PDOS in Fig.
3(a), the contribution from O 2$p$ PDOS to the feature A enhances
significantly making almost equal to that from Co 3$d$ state. It is
thus evident that Pr 4$f$ states play an important role in
determining the electronic states in the vicinity of $\epsilon_F$
and hence the electronic properties of this system.

We now study the effect of different spin configurations of
Co$^{3+}$ ion on the valence band calculated using GGA+$U$ method.
The Pr 4$f$, Co 3$d$ and O 2$p$ PDOS corresponding to Co$^{3+}$ ion
in LS, IS and HS configurations are plotted in Figs. 4(a), 4(b) and
4(c), respectively. It is evident from the figure that the
insulating gap of about 0.27 eV observed for LS configuration is not
present in the other cases. In addition, there are several changes
in the valence band corresponding to the IS and HS state compared to
that for the LS state. The low energy spread of the valence band
increases from LS to IS by about 1 eV and it further increases by
about 0.3 eV in the HS state. The Co 3$d$ character of the energy
bands at lower energies gradually enhances with the increase in spin
state. The Co 3$d$ character of the energy bands close to
$\epsilon_F$ observed in LS state reduces significantly in the IS
state. In the HS state, the energy band with dominant Co 3$d$
character appear below -5 eV.

In Fig. 5, we show the background subtracted experimental valence
band spectra of PrCoO$_{3}$ recorded at room temperature using
monochromatic Al K$\alpha$, He {\scriptsize II} and He {\scriptsize
I}. All the spectra show zero intensity at the Fermi level
indicating the insulating behavior of the compound. All the spectra
exhibit three distinct peaks at about 1.1, 3.2 and 4.9 eV marked by
A, B and C, respectively. The Al K$\alpha$ spectrum is dominated by
the intensity of peak A. This trend is reversed in the He
{\scriptsize II} and He {\scriptsize I} data. Due to the matrix
element effect, the photoemission cross section corresponding to
various electronic states changes significantly with the change in
photon energy. The relative cross section for O 2$p$ states with
respect to that of the Co 3$d$ and Pr 4$f$ states is significantly
higher in ultraviolet photon energies. The $x$-ray energies
correspond to the reverse case. Thus, large intensity of the
features B and C in the He {\scriptsize I} and He {\scriptsize II}
spectra compared to the intensity of the feature A indicates that
the features B and C has dominant O 2$p$ character and the feature A
is essentially contributed by the Co 3$d$ and Pr 4$f$ states. In
order to investigate the energy positions of the Pr 4$f$ and Co 3$d$
contributions, we compare the Al $K\alpha$ spectrum of PrCoO$_3$
with that of LaCoO$_3$. Clearly, the intensity around 1 eV has
dominant Co 3$d$ character. The large intensity difference observed
around 1.7 eV as marked by D in the figure correspond to the
intensities from Pr 4$f$ states. This is consistent with the
previous studies.\cite{pandey3,saitoh2}

A comparison of the results in Figs. 4 and 5 indicates that the
experimental results in Fig. 5 correspond closely to the results in
Fig. 4(b) indicating large IS state contributions at room
temperature. In order to bring out the comparison more clearly, we
show the spin-integrated DOS corresponding to IS state in Fig. 5.
Evidently, the character of various features and their energy
positions observed in the calculated DOS are very close to the
experimental results. The experimental O 2$p$ non-bonding band
appears at slightly higher binding energy compared to that observed
in the calculated results. Such small shift in energy position of
the O 2$p$ non-bonding states is often observed due to the neglect
of electron correlation effect among O 2$p$ electrons.\cite{sarma}
It is to note here that the observation of Co 3$d$ band at about 6
eV binding energy in the HS state has no resemblance with the
experimental spectra. This indicates that the contribution from HS
state at room temperature is essentially absent.

It is important to note here that although features in the DOS
corresponding to IS state provide a remarkable representation of the
experimental spectra, the insulating gap observed in the
experimental spectra as well as in the bulk measurements could not
be captured in our calculations for these parameters. One needs to
enhance the values to $U$ unrealistically high to achieve such
insulating phase. In that case, the features in the DOS will be
significantly different. It is well known that orbital ordering
plays an important role in determining the electronic structure of
these systems. To get an insulating state, one should consider the
orbital ordering in the calculations.\cite{korotin} However, such
considerations does not have significant influence in the energy
position of the features.

We now turn to the temperature evolution of the electronic
structure. In Fig. 6, we show the valence band spectra collected at
different temperatures. The XP spectra shown in Fig. 6(a) exhibit
almost identical lineshape of the spectra at 300 K and 150 K. The
band edge appears to shift slightly towards higher binding energies.
The He {\scriptsize II} spectra shown in Fig. 6(b), however, exhibit
significant change with temperatures. The decrease in temperature
leads to gradual increase in intensity of the feature around 5 eV
characterized as the bonding features. Subsequently, the intensity
of the features around 1 eV decreases. Since, the photoemission
cross section corresponding to O 2$p$ states is significantly
enhanced in the He {\scriptsize II} photon energies, the change in
Fig 6(b) suggests gradual enhancement of the O 2$p$ character of the
bonding feature with the decrease in temperature. We have expanded
the near $\epsilon_F$ region of the He {\scriptsize II} spectra in
Fig. 6(c) to clearly investigate the spectral changes. It is clear
that the intensity at about 1 eV reduces to a lower value at 250 K
compared to the intensity at 300 K. However, further reduction in
temperature does not have significant influence in the intensity of
the feature.

From the {\it ab initio} results (see Fig. 4), we have observed that
the increase in spin state leads to an enhancement of the Co 3$d$
character at higher binding energies and the O 2$p$ contributions
shift towards $\epsilon_F$. Thus, the increase in O 2$p$ character
at higher binding energies with the decrease in temperature
indicates that the contribution from LS state enhances gradually
with the decrease in temperature. These results thus provide a
direct experimental evidence of dominant IS contribution at 300 K
and an evolution towards LS state with the decrease in temperature.
We could not measure the spectra at lower temperatures due to the
charging effect.

In order to investigate the shift of the valence band edge leading
to an enhancement of the band gap with the decrease in temperatures,
we have extracted the Co 3$d$ and Pr 4$f$ part by subtracting the O
2$p$ contributions appearing at higher binding energies. The
subtraction procedure is shown in Fig. 6, where the lines in the
figure represent the O 2$p$ spectra obtained by using a combination
of Lorentzians convoluted by a Gaussian. The extracted features are
shown in Figs. 7(a) and (b) for XP and He {\scriptsize II} spectra,
respectively. Although this feature is contributed by Pr 4$f$ and Co
3$d$ states, the large change in photon energy (from 40.8 eV to
1486.6 eV) thereby different change in the photoemission cross
section does not have significant influence in the lineshape of the
feature. This suggests a strong overlap between the 4$f$ and 3$d$
contributions in this energy range.

We have expanded the band edge part of the high-resolution He
{\scriptsize II} spectra in Fig. 7(c). Interestingly, the band edge
gradually shifts towards higher binding energies with the decrease
in temperature. The shift is most significant between 300 K and 275
K. In the temperature range of 275 K to 200 K, the change is almost
negligible and further decrease in temperature again leads to a
shift of the band edge towards higher binding energies. Such change
can be correlated to the transport data too. In Fig. 7(d) we show
the activation gap calculated from the resistivity data.
Interestingly, the temperature induced change of the band gap
corresponds well to the shift of the band edge in Fig. 7(c).

In Fig. 8, we show the calculated majority and minority O 2$s$ and
Pr 5$p$ PDOS corresponding to LS, IS and HS states. These PDOS can
be divided into four regions as marked in the figure. In the LS
state, the region 1 spreading over -15.1 to -16.3 eV has dominating
Pr 5$p$ character and can be attributed to the spin-orbit split Pr
5$p_{3/2}$ states. The region 4 (between -18.8 and -19.3 eV) has
predominantly Pr 5$p$ character corresponding to spin-orbit split Pr
5$p_{1/2}$ state. In region 2 (from -16.4 to -17.1 eV), the
contribution from O 2$s$ states is most evident and region 3 (from
-17.4 to -18.0 eV) has highly mixed O 2$s$ and Pr 5$p$ characters.
It is evident from the figures that the overall shape and spread of
O 2$s$ and Pr 5$p$ bands remain almost the same for all the spin
state configurations. However, the energy position of these bands
are very much sensitive to the spin state of the Co ion. The whole
pattern is shifted to lower energy side by about about 1.0 and 1.1
eV in the IS and HS states, respectively.

The calculated O 2$s$ and Pr 5$p$ PDOS for the IS state are compared
with the experimental spectra in Fig. 9. There are two distinct
features in the experimental spectra centered around 16.8 and 20.6
eV. The intensity ratio of the two features are different from 1:2
expected for Pr 5$p$ signals due to the overlap of the O 2$s$ signal
appearing in this energy range. In order to identify the character
of the features, we overlap the Mg $K\alpha$ spectrum over the Al
$K\alpha$ spectrum. This change in photon energy leads to a larger
enhancement of the photoemission cross section of the O 2$s$ states
compared to that for Pr 5$p$ states. \cite{yeh} A normalization by
the Pr 5$p_{3/2}$ peak intensity exhibits an enhancement in the
energy region of 17 - 20 eV. This clearly indicates that the
presence of O 2$s$ feature in this energy region.

The calculated DOS corresponding to the IS state is shown in Fig.
9(b). Evidently, the calculated DOS provide a remarkable
representation of the experimental spectra in terms of energy
position and relative intensities. It is clear that the energy
positions obtained for IS and HS configurations are very similar to
to the experimental results and the energies for LS configuration is
very different.

Finally, we discuss the electronic occupancies of different orbitals
and corresponding magnetic moments obtained from self-consistent
calculations by considering initial spin state of Co ion in LS, IS
and HS configurations. The electronic occupancies of Pr 4$f$, Co
$t_{2g}$ and Co $e_{g}$ for majority and minority spin channels
corresponding to LS, IS and HS states are given in Table I. The
occupancies of Pr 4$f_{\uparrow}$ and Pr 4$f_{\downarrow}$ are
closer to 2.2 and 0.2, respectively, and insensitive to the spin
state of Co ion. However, the occupancies of $t_{2g\uparrow}$,
$t_{2g\downarrow}$, $e_{g\uparrow}$ and $e_{g\downarrow}$ are very
much sensitive to initial spin state configurations as expected. The
total occupancy of Co 3$d$ orbitals is significantly higher
($\sim$~6.69) than 6 expected for Co$^{3+}$ and gradually decreases
with the increase in spin state as also observed in the case of
LaCoO$_{3}$ compounds in earlier studies.\cite{saitoh1, pandey1}
Interestingly, $e_{g\downarrow}$ orbitals are also partially
occupied; 0.56, 0.44 and 0.42 electrons in the results corresponding
to LS, IS and HS states, respectively. This is not {\it a priori}
expected for Co$^{3+}$ electronic configuration. Moreover, one can
also see a remarkable increase in the total occupancies of Pr 4$f$
(by $\sim$0.4). This increment in the occupancies is a clear
indication of the strong hybridization between Pr 4$f$ with the
valence electrons having O 2$p$ and Co 3$d$ characters facilitating
large charge transfer. This is also manifested in Fig. 3; the shape
of the Co 3$d$ and O 2$p$ PDOS in the energy range of Pr 4$f$ bands
are very similar to Pr 4$f$ PDOS.

Spin and orbital part of the magnetic moment of Pr 4$f$, Co 3$d$ and
O 2$p$ for different spin state is given in Table II. Spin part of
the magnetic moment for Pr 4$f$ state is almost insensitive to the
spin state of Co ion and it is closer to 2 $\mu_{B}$. This indicates
that Pr$^{3+}$ ion is in spin triplet state. Spin part of the
magnetic moments for Co 3$d$ state when Co ion is in LS, IS and HS
states are 0.03, 1.79 and 2.96 $\mu_{B}$, respectively.
Surprisingly, the orbital part of the magnetic moment for Pr 4$f$
comes out to be negative and strongly depends on the spin state of
the Co ion. It's magnitude increases with the increase in spin
state. The negative sign indicates that the orbital and spin moments
are antiparallely coupled. The orbital moment of the Co 3$d$
electrons is almost zero for LS and IS state and for HS state it is
0.21 $\mu_{B}$. Interestingly, one can also see the contribution of
spin part of magnetic moment from O 2$p$ electrons when Co ion is in
HS state.

Using the values of magnetic moments given in Table II, one can
calculate the effective magnetic moment of PrCoO$_{3}$. This
quantity can be directly compared with susceptibility data. The
calculated values obtained for LS, IS and HS states are about 2.0,
3.1 and 4.3 $\mu_{B}$, respectively. The value of the effective
magnetic moments thus obtained for the IS state is closer to the
experimental value of about 3.6 $\mu_{B}$ obtained by fitting
magnetic susceptibility data between 100-150 K by using Curie-Weiss
law.\cite{tsubouchi}

\section{Conclusions}

In summary, we have investigated the detailed electronic structure
of PrCoO$_3$ and its temperature evolutions using various forms of
{\em ab initio} calculations and high resolution photoemission
spectroscopy. We observe that GGA+$U$ calculations provide a good
description of the ground state properties. Partial density of
states obtained for various configurations and parameters exhibit
signature of strong hybridization of Pr 4$f$ states with Co 3$d$ and
O 2$p$ valence electrons. In addition to the electron correlations,
spin-orbit coupling plays an important role in determining the
electronic structure of this compound.

The calculated Pr 4$f$, Pr 5$p$, Co 3$d$, O 2$s$ and O 2$p$ partial
density of states corresponding to intermediate spin (IS) state of
Co ion provide a good representation of the experimental spectra at
300 K. The contribution from high spin state configurations appears
to be negligible. Pr 5$p$ and O 2$s$ core level spectra could be
captured reasonably well within this formalism. The decrease in
temperature from 300 K leads to an enhancement of the LS
contributions in the electronic structure. The band gap gradually
enhances with the decrease in temperature as observed in the
transport data.

The calculated effective magnetic moment for IS state is also
consistent with the magnetic susceptibility data. The calculated
spin moment of Pr 4$f$ electrons indicates that the Pr$^{3+}$ ion is
in spin triplet state and the spin-orbit coupling is antiparallel.
The orbital part of the magnetic moment of Pr ion is highly
sensitive to the spin state of the Co ion. Its magnitude increases
gradually with the increase in spin state.

\section{Acknowledgements}

SP is thankful to CSIR, India, for financial support.


\pagebreak

\section{Figure Captions:}

Fig. 1: (color online) O 1$s$ core level spectra recorded at 300 K
(solid circles) and 150 K (hollow circles). The shaded region shows
the impurity contributions.

Fig. 2: (color online) (a) O 2$p$ (dashed line) and Co 3$d$ partial
density of states having $t_{2g}$ (thin solid line) and $e_g$ (thick
solid line) symmetries and (b) Pr 4$f$ partial density of states
obtained from GGA calculation. (c) O 2$p$ (dashed line) and Co 3$d$
partial density of states having $t_{2g}$ (thin solid line) and
$e_g$ (thick solid line) symmetries and (d) Pr 4$f$ partial density
of states obtained from GGA calculation including spin-orbit
coupling (SOC).

Fig. 3: (color online) (a) O 2$p$ (dashed line) and Co 3$d$ partial
density of states having $t_{2g}$ (thin solid line) and $e_g$ (thick
solid line) symmetries and (b) Pr 4$f$ partial density of states
obtained from GGA+\emph{U} calculation when on-site Coulomb
correlation between Co 3$d$ electrons is considered. (c) O 2$p$
(dashed line) and Co 3$d$ partial density of states having $t_{2g}$
(thin solid line) and $e_g$ (thick solid line) symmetries and (d) Pr
4$f$ partial density of states obtained from GGA calculation when
on-site Coulomb correlation between Co 3$d$ and Pr 4$f$ electrons is
considered. In all the calculations spin-orbit coupling (SOC) is
included.

Fig. 4: (color online) Calculated Pr 4$f$ (thin solid line), Co 3$d$
(thick solid line) and O 2$p$ (dashed line) partial density of
states corresponding to (a) low spin (LS), (b) intermediate spin
(IS) and (c) high spin (HS) configurations using GGA+$U$ method
including spin-orbit coupling.

Fig. 5: (color online) Experimental valence band spectra collected
at room temperature using Al $K\alpha$, He {\scriptsize II}  and He
{\scriptsize I} radiations. The lines denote the calculated Pr 4$f$
(thin solid line), Co 3$d$ (thick solid line) and O 2$p$ (dashed
line) partial density of states corresponding to intermediate spin
configuration of Co ion obtained from GGA+\emph{U} calculation
including spin-orbit coupling. Room temperature experimental valence
band spectrum of LaCoO$_{3}$ recorded using Al $K\alpha$ radiation
is also shown by hollow circles.

Fig. 6: (color online) (a) Valence band spectra collected at 300 K
(hollow circles) and 150 K (plus signs) using Al $K\alpha$
radiation. (b)  Valence band spectra collected at 300 K (hollow
circles), 250 K (solid circles), 200 K (solid triangles) and 150 K
(plus signs) using He {\scriptsize II} radiation. (c) Spectra
collected at 300 K (hollow circles), 250 K (solid circles), 200 K
(solid triangles) and 150 K (plus signs) in the narrow region of the
Fermi level using He {\scriptsize II} radiation. Solid lines in
panels (a) and (b) indicate the contribution from O 2$p$ states.

Fig. 7: (color online) Co 3$d$ and Pr 4$f$ contributions extracted
(a) from the XP spectra at 300 K (solid triangles) and 150 K (plus
signs), and (b) from the He {\scriptsize II} spectra at 300 K (solid
triangles), 225 K (hollow circles) and 150 K (plus signs).
Temperature evolution of the spectra in the vicinity of Fermi level
is shown in panel (c) and band gap obtained from resistivity data is
shown in panel (d).

Fig, 8: (color online) Pr 5$p$ (solid line) and O 2$s$ (dashed line)
PDOS calculated using (a) low spin (LS), (b) intermediate spin (IS)
and (c) high spin (HS) configurations of Co ion.

Fig. 9: (color online) (a) Background subtracted La 5$p$ and O 2$s$
core level spectra collected at 300 K using Al $K\alpha$ (hollow
circles) and Mg $K\alpha$ (solid lines). (b) Calculated Pr 5$p$
(solid lines) and O 2$s$ (dashed lines) PDOS corresponding to
intermediate spin (IS) state configuration.

\pagebreak


\begin{table}
\vspace{6ex}
 \caption{ Electronic occupancies of Pr 4$f$,
Co $t_{2g}$ and Co $e_{g}$ orbitals obtained from GGA+$U$
calculations for low spin (LS), intermediate spin (IS) and high spin
(HS) configurations of Co ion in PrCoO$_3$. }
 \vspace{2ex}
\begin{ruledtabular}
\begin{tabular}{|c|c|c|c|c|c|c|c|c|}


 & Pr 4$f\uparrow$ & Pr 4$f\downarrow$ & Pr 4$f$ & $t_{2g\uparrow}$
&
$e_{g\uparrow}$ & $t_{2g\downarrow}$ & $e_{g\downarrow}$ & Co 3$d$ \\
\hline
 LS & 2.18 & 0.21 & 2.39& 2.77 & 0.59 & 2.77 & 0.56 & 6.69
\\
 \hline
 IS & 2.16 & 0.22 & 2.38 & 2.8 & 1.4 & 1.98 & 0.44 & 6.62 \\
   \hline
 HS & 2.17 & 0.22 & 2.39 & 2.83 & 1.89 & 1.36 & 0.42 & 6.5 \\

\end{tabular}
\end{ruledtabular}
\end{table}
\vspace{8ex}

\begin{table}
\vspace{12ex}
 \caption{Spin and orbital part of magnetic
moments of Pr 4$f$, Co 3$d$ and O 2$p$ orbitals when Co ion is in
low spin (LS), intermediate spin (IS) and high spin (HS)
configurations.}
 \vspace{2ex}
\begin{ruledtabular}
\begin{tabular}{|c|c|c|c|c|c|c|c|}
 & \multicolumn{2}{c|}{Pr 4$f$ ($\mu_B$)} & \multicolumn{2}{c|}{Co
3$d$ ($\mu_B$)} & \multicolumn{2}{c|}{O 2$p$ ($\mu_B$)} & $\mu_{eff}$ ($\mu_B$)\\
\hline
 & Spin & Orbital & Spin & Orbital & Spin & Orbital & \\ \hline
LS & 1.97 & -0.47 & 0.03 & $\sim$ 0 & 0.02 & $\sim$ 0 & 2.0 \\
\hline
 IS & 1.96 & -0.74 & 1.79 & 0.03 & 0.02 & 0.005 & 3.1 \\
\hline
 HS & 1.99 & -0.95 & 2.96 & 0.21 & 0.13 & 0.005 & 4.3 \\


\end{tabular}
\end{ruledtabular}
\end{table}

\end{document}